\documentclass[twocolumn,preprintnumbers,floatfix,amsmath,amssymb,superscriptaddress]{revtex4} 

\usepackage{graphicx}
\usepackage{dcolumn}
\usepackage{bm}

\begin{document} 

\setlength{\topmargin}{0in}

\title{Dual structures in (1-$x$)Pb(Zn$_{1/3}$Nb$_{2/3}$)O$_3$-$x$PbTiO$_3$ 
ferroelectric relaxors}
\author{Guangyong Xu}
\affiliation{Physics Department, Brookhaven National Laboratory, Upton, 
New York 11973}
\author{H. Hiraka} 
\affiliation{Physics Department, Brookhaven National Laboratory, Upton, 
New York 11973}
\affiliation{Institute for Material Research, Tohoku University, 
Sendai 980-8577, Japan}
\author{K. Ohwada}
\affiliation{Synchrotron Radiation Research Center, JAERI, Sayo-gun Hyogo 
679-5148, Japan}
\author{G. Shirane} 
\affiliation{Physics Department, Brookhaven National Laboratory, Upton, 
New York 11973}
\date{\today} 
 
\begin{abstract} 

We performed x-ray diffraction studies on a series of 
(1-$x$)Pb(Zn$_{1/3}$Nb$_{2/3}$)O$_3$-$x$PbTiO$_3$ (PZN-$x$PT) single crystals
with different incident photon energies, and therefore different penetration 
depths. Our results show that outer-layers of $\sim 10$ to $50$~$\mu$m 
thick are present 
in all samples. The structure of those outer-layers  
is different from that of the inside of the crystals, by having much greater
(rhombohedral) distortions. 
With increasing $x$, rhombohedral-type lattice distortions develop, both in 
the outer-layer and the inside.

\end{abstract}

\maketitle 

\begin{figure}[ht]
\includegraphics[width=\linewidth]{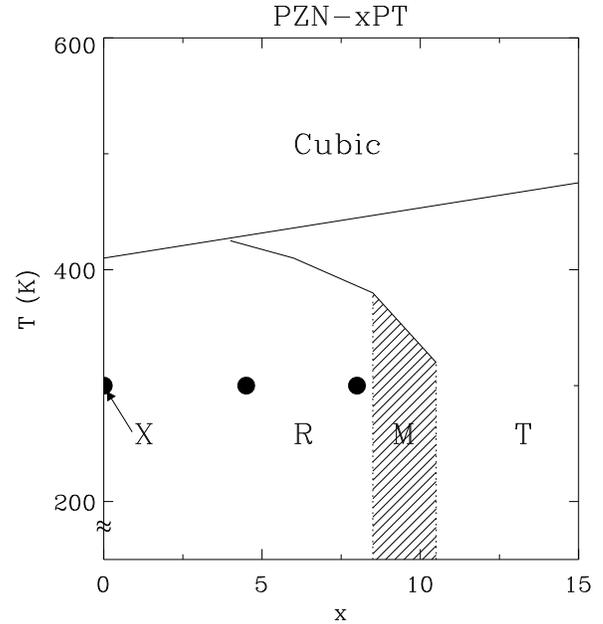}
\caption{Schematic of the revised phase diagram of PZN-$x$PT in zero field. 
The dots show the PT concentrations of the crystals in our measurements.}
\label{fig:1}
\end{figure}

The relaxor system Pb(Zn$_{1/3}$Nb$_{2/3}$)O$_3$ (PZN) and its solid solutions 
with PbTiO$_3$ ($x$PT) have attracted much attention recently due to their 
extraordinary piezoelectric properties~\cite{PZT1}, and potential  
industrial application as high-performance actuators. The high piezoelectric
response is shown to be directly related to a particular region in the phase 
diagram~\cite{PZN_phase1,PZN_phase2,Universal_phase,PZN_phase,Uesu}. 
In zero external electric field, for temperatures below the Curie 
Temperature $T_C$,
the system has long been believed to be in a rhombohedral (R) phase for 
small $x$, and tetragonal (T) for large $x$, separated by a narrow region of 
monoclinic (M) phase~\cite{Universal_phase,PZN_phase,Uesu} (
see Fig.~\ref{fig:1}). 
With the application of an external electric field along the [001] 
direction, however, the M phase can be induced in PZN-$x$PT systems with
smaller $x$ values (the zero filed rhombohedral region). Even with the removal 
of the field, the M phase still remains~\cite{PZN_field,PZN_efield}.

Only recently has the structure of pure PZN been well determined. 
It was believed to be in a so-called microdomain
state below $T_C=410$~K~\cite{Mulvihill}, 
which only transforms into
a macrodomain state with external electric field poling along [111] direction.
Lebon {\it et al.}~\cite{Lebon} reported the first explicit zero field 
structural measurements using 
conventional x-ray diffraction (Cu $K_\alpha$ 8.9~keV). They observed
the rhombohedral distortion for temperatures below $T_C$ as 
previously believed. 
However, more recently,  high energy x-ray (67~keV) diffraction work, which 
penetrates more deeply into the sample,  by Xu 
{\it et al.}~\cite{PZN_Xu} on pure PZN single 
crystals shows that the inside of the crystal does not have any measurable 
(rhombohedral) lattice distortion. This new phase X 
has an average cubic {\it lattice} ($a=b=c, \alpha=\beta=\gamma=90^\circ$), yet
the true symmetry is still unknown. 
Another important finding during the x-ray diffraction work on PZN was the 
existence of an ``outer-layer'' with different structures than the inside 
of the crystal. Lower energy x-rays (10.2~keV) were used to study the PZN 
single crystal, and a rhombohedral distortion was found to exist below $T_C$ 
for the outer most $\sim10$ to $50$~$\mu$m of the crystal~\cite{PZN_Xu}. 

In this letter, we present 
x-ray diffraction result using both 67 and 10.2~keV x-rays on PZN, 
PZN-4.5\%PT and PZN-8\%PT single crystals, measured at 300~K, below $T_C$ 
for all three systems.  Our results show that
the inside of both 4.5PT and 8PT crystal has rhombohedral distortions, 
unlike the phase X in pure PZN. On the other hand, all three systems show
clear difference between the inside and outer-layer structures.

The PZN single crystal  was grown at the Simon Fraser 
University in Canada (the same
crystal used in Ref~\cite{PZN_Xu}), and one of the 8PT crystal was grown 
at the Pennsylvania State University, previously studied by Ohwada 
{\it et al.}~\cite{PZN_efield}. We  have also used
4.5PT and 8PT single crystals provided by H.~C. Materials.
All samples have been repeatedly heated well above 
$T_C$ and cooled down to eliminate any residue electric field polling 
effect. The 67~keV x-ray measurements were performed at X17B1 beamline 
of the National Synchrotron Light Source (NSLS). All the 67~keV x-ray 
diffraction measurements were performed in transmission mode so that the 
inside of the bulk crystal was probed (penetration depth $\sim400$~$\mu$m).
The 10.2~keV x-ray measurements were performed in reflection mode 
with penetration depth $\sim10$~$\mu$m, at X22A beamline of NSLS. 


\begin{figure}[ht]
\includegraphics[width=\linewidth]{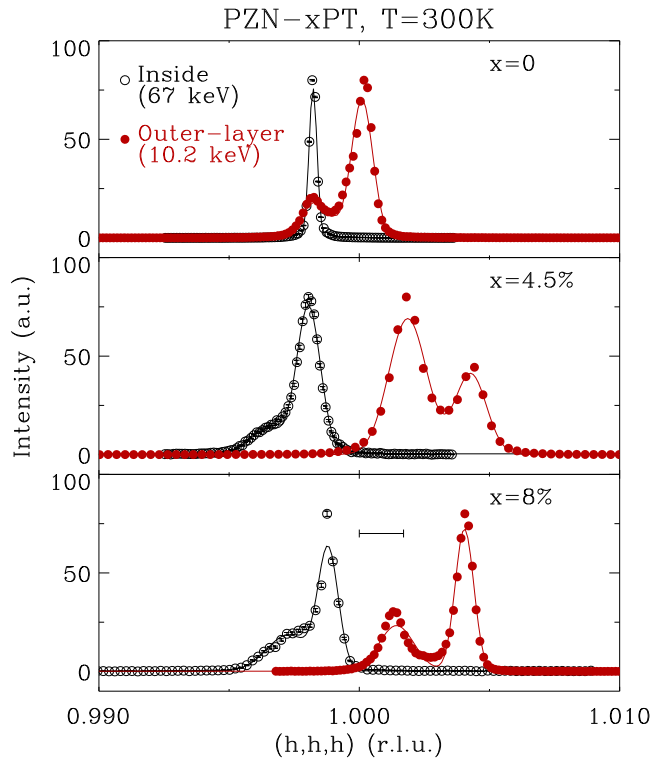}
\caption{Longitudinal x-ray diffraction profiles  about the (111) 
Bragg peak of PZN, PZN-4.5\%PT, and PZN-8\%PT single crystals. 
The x-ray energy was 67~keV for measuring the inside structure
and 10.2~keV for measuring the outer-layer. The horizontal bar shows the 
estimated systematic error between the two sets of measurement, from
wavelength and $2\theta_0$ calibrations. }
\label{fig:2}
\end{figure}

In Fig.~\ref{fig:2}, we show x-ray diffraction profiles obtained by doing 
longitudinal scans ($\theta-2\theta$ scans) about the (111) Bragg peak on 
the $x=0$, $4.5\%$, and $8\%$ samples. The units are multiples of the 
reciprocal lattice units (r.l.u.) $a^*=2\pi/4.060$~\AA~$=1.5476$~\AA$^{-1}$.
For pure PZN, with 67~keV x-rays, the Bragg profile is one 
single sharp peak without any splittings, as expected from an 
undistorted lattice (phase X)~\cite{PZN_Xu}. The inside of the 4.5PT and 8PT
crystals, however, are quite different. Both have a main
peak and a ``shoulder'' on the left side, indicating that the lattice is
(rhombohedrally) distorted.

Similar measurements on the same samples were performed with 10.2~keV x-rays.
The results are also plotted in Fig.~\ref{fig:2}. The outer-layer Bragg 
profiles of all three samples have two peaks. For pure PZN, it is 
clearly different from the undistorted structure of the inside of the crystal. 
Based on the penetration length of 10.2~keV x-rays ($\xi\sim 13$~$\mu$m), 
we believe the depth of this rhombohedrally distorted region is about $\sim10$ 
to $50$~$\mu$m. In addition, the peaks are shifted to higher 
$q$ positions for all three samples, compared to the 67~keV x-ray results.

\begin{figure}[ht]
\includegraphics[width=\linewidth]{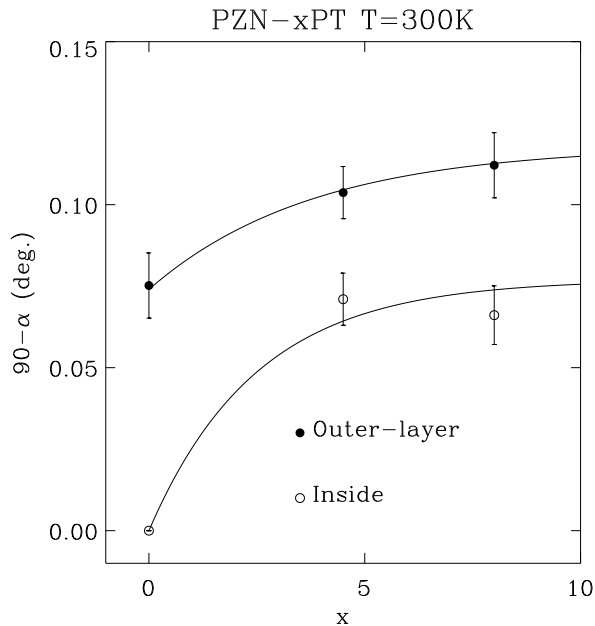}
\caption{Rhombohedral angle 90-$\alpha$, for the inside and outer-layer 
structures of PZN-$x$PT ($x=0$, $4.5\%$, and $8\%$).}
\label{fig:3}
\end{figure}

For a rhombohedral lattice, the \{111\} peaks split into two because of two 
different $q_{(111)}$ lengths associated with different domains.
With a fit to two Gaussians 
(solid line in Fig.~\ref{fig:2}), we can extract the positions of the two 
peaks and therefore determine the rhombohedral lattice parameter $a$ 
and rhombohedral angle $\alpha$, as shown in Table.~\ref{tab:1} and 
Fig.~\ref{fig:3}.

\begin{table}
\caption{Rhombohedral lattice parameters for the inside and 
outer-layer structures of PZN-$x$PT single crystals.}
\begin{ruledtabular}
\begin{tabular}{cccc}
&  & $a$ (\AA) & $\alpha$ (deg.) \\
\hline
$x=0$ & inside & 4.067(1) & -\\
$x=0$ & outer-layer&4.061(1)& 89.92\\
$x=4.5\%$ & inside & 4.070(1) &89.93\\ 
$x=4.5\%$ & outer-layer & 4.045(1)& 89.90\\
$x=8\%$ & inside & 4.066(1)& 89.93\\
$x=8\%$ &  outer-layer& 4.046(1)& 89.89

\end{tabular}
\end{ruledtabular}
\label{tab:1}
\end{table}

Here we can see that for pure PZN, the outer-layer structure and the inside
structure are distinctly different (undistorted {\it vs.} rhombohedrally 
distorted). For the 4.5PT and 8PT crystals, both the outer-layer and inside
are rhombohedrally distorted, yet the distortion and the lattice 
parameter are still quantitatively different.
The rhombohedral angle $\alpha$, which is a direct measure of how much the 
lattice is distorted, is also shown in Fig.~\ref{fig:3}.
The rhombohedral distortions of the outer-layers in our measurements are 
consistent with results from previous studies on pure PZN~\cite{Lebon}, 
PZN-4.5\%PT~\cite{PZN_Forrester} and PZN-8\%PT~\cite{Polarization}, which
in fact did measure the outer-layer structure, due to the limited penetration 
length of the x-ray energy used and/or the powder nature of the samples.
It is very important to note that for same $x$ values, 
the outer-layer is always  more distorted than the inside. In addition,
the rhombohedral distortion develops in both the outer-layer and the inside
with increasing $x$.

Similar behavior has also been observed in another prototype lead-oxide 
perovskite system Pb(Mg$_{1/3}$Nb$_{2/3}$)O$_3$-$x$PbTiO$_3$ (PMN-$x$PT).
High $q$-resolution neutron scattering work by Gehring 
{\it et al.}~\cite{PMN-10PT} and Xu {\it et al.}~\cite{Xu1}
show that PMN-10\%PT and PMN-20\%PT also transform into phase X instead of 
R below $T_C$. However, in PMN-27\%PT~\cite{Xu1}, 
the low temperature phase was found to be rhombohedral instead of X. These 
results indicate that the ground state for the inside of PMN-$x$PT is similar
to that of PZN-$x$PT system, i.e., phase X for small $x$ values, 
and with increasing
$x$, rhombohedral distortions develop and the system changes into a 
true rhombohedral phase. 
On the other hand, even for very small $x$ values in PMN-$x$PT, 
rhombohedral distortion has been observed by conventional x-ray diffraction 
measurements~\cite{PMN_Ye}, which probes the outer-layer. It shows that the 
difference between the inside and outer-layer structures is also present 
in PMN-$x$PT series.

What is the driving force behind these differences between the outer-layer and
the inside of these crystals? Why is the inside of pure PZN (phase X) 
undistorted? 
These are questions yet to be answered. In our current picture, the
polar nanoregions (PNR) play a very important role in answering these 
questions. 

The concept of PNR was first introduced by Burns and 
Dacol~\cite{Burns} in their attempt to interpret the optical index of
refraction data on several ferroelectric and relaxor compounds, including 
both PZN and PMN. It was then widely used in describing the common properties
of relaxor systems~\cite{Smolensky,Cross}. The PNR start to appear at the Burns
temperature $T_d$, which is a few hundred degrees above  
$T_C$, and grow with cooling. Lately, Hirota {\it et al.}~\cite{PMN_diffuse} 
proposed that the PNR are not only polarized regions, but also displaced from 
their surrounding lattices along the polarization direction 
(``uniform phase shift'').

In pure PZN, although the inside of the crystal has an undistorted lattice,
ferroelectric soft TO phonon measurements~\cite{Stock1} still show signatures
of global ferroelectric polarization below $T_C$. It is usually true 
that ferroelectric polarization would lead to lattice distortions. However,
in phase X, the lattice distortion is not achieved, possibly due to the energy
barrier created by the shifted PNR. Phase X is therefore a special phase in 
which the lattice is polarized yet not distorted. A balance is established 
inside the lattice between the polarization, which wants to distort the lattice;
and the energy barrier, coming from phase shifted PNR, which prevents a global
lattice distortion to form. A small driving force,
such as an external electric field, would be able to ``pole'' the crystal
and create a global rhombohedral lattice. Another possible 
driving force, is the boundary conditions near the surface of 
the crystal. In pure PZN, the outer-layer is rhombohedrally distorted, yet 
the inside is undistorted. In 4.5PT and 8PT, the rhombohedral distortion in 
the outer-layer is more predominant than that inside the crystal. 
Moving from the inside to the outer-layer, the balance between the 
ferroelectric polarization and the energy barrier from the PNR is also tuned
in favor of the distortion.

We believe the existence of an outer-layer in the relaxor system PZN-$x$PT, and 
very likely in other relaxor systems such as PMN-$x$PT, is extremely interesting
and helps explaining many discrepancies between previous conventional x-ray 
powder diffraction and current neutron scattering and high energy x-ray 
diffraction results. More work has been carried out on detailed temperature
dependent structure analysis of the outer-layer and inside of the PZN-$x$PT
series, and will be discussed elsewhere.

We would like to thank P.~Gehring, J.~Hill, B.~Ocko, S.~Shapiro, C.~Stock 
and Z.~Zhong for stimulating discussions.
Financial support from the U.S. Department of Energy under contract 
No.~DE-AC02-98CH10886 is also gratefully acknowledged.


\begin{thebibliography}{21}
\expandafter\ifx\csname natexlab\endcsname\relax\def\natexlab#1{#1}\fi
\expandafter\ifx\csname bibnamefont\endcsname\relax
  \def\bibnamefont#1{#1}\fi
\expandafter\ifx\csname bibfnamefont\endcsname\relax
  \def\bibfnamefont#1{#1}\fi
\expandafter\ifx\csname citenamefont\endcsname\relax
  \def\citenamefont#1{#1}\fi
\expandafter\ifx\csname url\endcsname\relax
  \def\url#1{\texttt{#1}}\fi
\expandafter\ifx\csname urlprefix\endcsname\relax\def\urlprefix{URL }\fi
\providecommand{\bibinfo}[2]{#2}
\providecommand{\eprint}[2][]{\url{#2}}

\bibitem[{\citenamefont{Park and Shrout}(1997)}]{PZT1}
\bibinfo{author}{\bibfnamefont{S.-E.} \bibnamefont{Park}} \bibnamefont{and}
  \bibinfo{author}{\bibfnamefont{T.~R.} \bibnamefont{Shrout}},
  \bibinfo{journal}{J. Appl. Phys.} \textbf{\bibinfo{volume}{82}},
  \bibinfo{pages}{1804} (\bibinfo{year}{1997}).

\bibitem[{\citenamefont{Kuwata et~al.}(1982)\citenamefont{Kuwata, Uchino, and
  Nomura}}]{PZN_phase1}
\bibinfo{author}{\bibfnamefont{J.}~\bibnamefont{Kuwata}},
  \bibinfo{author}{\bibfnamefont{K.}~\bibnamefont{Uchino}}, \bibnamefont{and}
  \bibinfo{author}{\bibfnamefont{S.}~\bibnamefont{Nomura}},
  \bibinfo{journal}{Jpn. J. Appl. Phys.} \textbf{\bibinfo{volume}{21}},
  \bibinfo{pages}{1298} (\bibinfo{year}{1982}).

\bibitem[{\citenamefont{Kuwata et~al.}(1981)\citenamefont{Kuwata, Uchino, and
  Nomura}}]{PZN_phase2}
\bibinfo{author}{\bibfnamefont{J.}~\bibnamefont{Kuwata}},
  \bibinfo{author}{\bibfnamefont{K.}~\bibnamefont{Uchino}}, \bibnamefont{and}
  \bibinfo{author}{\bibfnamefont{S.}~\bibnamefont{Nomura}},
  \bibinfo{journal}{Ferroelectrics} \textbf{\bibinfo{volume}{37}},
  \bibinfo{pages}{579} (\bibinfo{year}{1981}).

\bibitem[{\citenamefont{Cox et~al.}(2001)\citenamefont{Cox, Noheda, Shirane,
  Uesu, Fujishiro, and Yamada}}]{Universal_phase}
\bibinfo{author}{\bibfnamefont{D.~E.} \bibnamefont{Cox}},
  \bibinfo{author}{\bibfnamefont{B.}~\bibnamefont{Noheda}},
  \bibinfo{author}{\bibfnamefont{G.}~\bibnamefont{Shirane}},
  \bibinfo{author}{\bibfnamefont{Y.}~\bibnamefont{Uesu}},
  \bibinfo{author}{\bibfnamefont{K.}~\bibnamefont{Fujishiro}},
  \bibnamefont{and} \bibinfo{author}{\bibfnamefont{Y.}~\bibnamefont{Yamada}},
  \bibinfo{journal}{Appl. Phys. Lett.} \textbf{\bibinfo{volume}{79}},
  \bibinfo{pages}{400} (\bibinfo{year}{2001}).

\bibitem[{\citenamefont{La-Orauttapong
  et~al.}(2002)\citenamefont{La-Orauttapong, Noheda, Ye, Gehring, Toulouse,
  Cox, and Shirane}}]{PZN_phase}
\bibinfo{author}{\bibfnamefont{D.}~\bibnamefont{La-Orauttapong}},
  \bibinfo{author}{\bibfnamefont{B.}~\bibnamefont{Noheda}},
  \bibinfo{author}{\bibfnamefont{Z.-G.} \bibnamefont{Ye}},
  \bibinfo{author}{\bibfnamefont{P.~M.} \bibnamefont{Gehring}},
  \bibinfo{author}{\bibfnamefont{J.}~\bibnamefont{Toulouse}},
  \bibinfo{author}{\bibfnamefont{D.~E.} \bibnamefont{Cox}}, \bibnamefont{and}
  \bibinfo{author}{\bibfnamefont{G.}~\bibnamefont{Shirane}},
  \bibinfo{journal}{Phys. Rev. B} \textbf{\bibinfo{volume}{65}},
  \bibinfo{pages}{144101} (\bibinfo{year}{2002}).

\bibitem[{\citenamefont{Uesu et~al.}(2002)\citenamefont{Uesu, Matsuda, Yamada,
  Fujishiro, Cox, Noheda, and Shirane}}]{Uesu}
\bibinfo{author}{\bibfnamefont{Y.}~\bibnamefont{Uesu}},
  \bibinfo{author}{\bibfnamefont{M.}~\bibnamefont{Matsuda}},
  \bibinfo{author}{\bibfnamefont{Y.}~\bibnamefont{Yamada}},
  \bibinfo{author}{\bibfnamefont{K.}~\bibnamefont{Fujishiro}},
  \bibinfo{author}{\bibfnamefont{D.~E.} \bibnamefont{Cox}},
  \bibinfo{author}{\bibfnamefont{B.}~\bibnamefont{Noheda}}, \bibnamefont{and}
  \bibinfo{author}{\bibfnamefont{G.}~\bibnamefont{Shirane}},
  \bibinfo{journal}{J. Phys. Soc. Jpn.} \textbf{\bibinfo{volume}{71}},
  \bibinfo{pages}{960} (\bibinfo{year}{2002}).

\bibitem[{\citenamefont{Noheda et~al.}(2002)\citenamefont{Noheda, Zhong, Cox,
  Shirane, Park, and Rehrig}}]{PZN_field}
\bibinfo{author}{\bibfnamefont{B.}~\bibnamefont{Noheda}},
  \bibinfo{author}{\bibfnamefont{Z.}~\bibnamefont{Zhong}},
  \bibinfo{author}{\bibfnamefont{D.~E.} \bibnamefont{Cox}},
  \bibinfo{author}{\bibfnamefont{G.}~\bibnamefont{Shirane}},
  \bibinfo{author}{\bibfnamefont{S.-E.} \bibnamefont{Park}}, \bibnamefont{and}
  \bibinfo{author}{\bibfnamefont{P.}~\bibnamefont{Rehrig}},
  \bibinfo{journal}{Phys. Rev. B} \textbf{\bibinfo{volume}{65}},
  \bibinfo{pages}{224101} (\bibinfo{year}{2002}).

\bibitem[{\citenamefont{Ohwada et~al.}(2003)\citenamefont{Ohwada, Hirota,
  Rehrig, Fujii, and Shirane}}]{PZN_efield}
\bibinfo{author}{\bibfnamefont{K.}~\bibnamefont{Ohwada}},
  \bibinfo{author}{\bibfnamefont{K.}~\bibnamefont{Hirota}},
  \bibinfo{author}{\bibfnamefont{P.~W.} \bibnamefont{Rehrig}},
  \bibinfo{author}{\bibfnamefont{Y.}~\bibnamefont{Fujii}}, \bibnamefont{and}
  \bibinfo{author}{\bibfnamefont{G.}~\bibnamefont{Shirane}},
  \bibinfo{journal}{Phys. Rev. B} \textbf{\bibinfo{volume}{67}},
  \bibinfo{pages}{094111} (\bibinfo{year}{2003}).

\bibitem[{\citenamefont{Mulvihill et~al.}(1997)\citenamefont{Mulvihill, Cross,
  Cao, and Uchino}}]{Mulvihill}
\bibinfo{author}{\bibfnamefont{M.~L.} \bibnamefont{Mulvihill}},
  \bibinfo{author}{\bibfnamefont{L.~E.} \bibnamefont{Cross}},
  \bibinfo{author}{\bibfnamefont{W.}~\bibnamefont{Cao}}, \bibnamefont{and}
  \bibinfo{author}{\bibfnamefont{K.}~\bibnamefont{Uchino}},
  \bibinfo{journal}{J. Am. Ceram. Soc} \textbf{\bibinfo{volume}{80}},
  \bibinfo{pages}{1462} (\bibinfo{year}{1997}).

\bibitem[{\citenamefont{Lebon et~al.}(2002)\citenamefont{Lebon, Dammak,
  Calvarin, and Ahmedou}}]{Lebon}
\bibinfo{author}{\bibfnamefont{A.}~\bibnamefont{Lebon}},
  \bibinfo{author}{\bibfnamefont{H.}~\bibnamefont{Dammak}},
  \bibinfo{author}{\bibfnamefont{G.}~\bibnamefont{Calvarin}}, \bibnamefont{and}
  \bibinfo{author}{\bibfnamefont{I.~O.} \bibnamefont{Ahmedou}},
  \bibinfo{journal}{J. Phys.: Condens. Matter} \textbf{\bibinfo{volume}{14}},
  \bibinfo{pages}{7035} (\bibinfo{year}{2002}).

\bibitem[{\citenamefont{Xu et~al.}(2003{\natexlab{a}})\citenamefont{Xu, Zhong,
  Bing, Ye, Stock, and Shirane}}]{PZN_Xu}
\bibinfo{author}{\bibfnamefont{G.}~\bibnamefont{Xu}},
  \bibinfo{author}{\bibfnamefont{Z.}~\bibnamefont{Zhong}},
  \bibinfo{author}{\bibfnamefont{Y.}~\bibnamefont{Bing}},
  \bibinfo{author}{\bibfnamefont{Z.-G.} \bibnamefont{Ye}},
  \bibinfo{author}{\bibfnamefont{C.}~\bibnamefont{Stock}}, \bibnamefont{and}
  \bibinfo{author}{\bibfnamefont{G.}~\bibnamefont{Shirane}},
  \bibinfo{journal}{Phys. Rev. B} \textbf{\bibinfo{volume}{67}},
  \bibinfo{pages}{104102} (\bibinfo{year}{2003}{\natexlab{a}}).

\bibitem[{\citenamefont{Forrester et~al.}(2001)\citenamefont{Forrester, Piltz,
  Kisi, and McIntyre}}]{PZN_Forrester}
\bibinfo{author}{\bibfnamefont{J.~S.} \bibnamefont{Forrester}},
  \bibinfo{author}{\bibfnamefont{R.~O.} \bibnamefont{Piltz}},
  \bibinfo{author}{\bibfnamefont{E.~H.} \bibnamefont{Kisi}}, \bibnamefont{and}
  \bibinfo{author}{\bibfnamefont{G.~J.} \bibnamefont{McIntyre}},
  \bibinfo{journal}{J. Phys.: Condens. Matter} \textbf{\bibinfo{volume}{13}},
  \bibinfo{pages}{L825} (\bibinfo{year}{2001}).

\bibitem[{\citenamefont{Noheda et~al.}(2001)\citenamefont{Noheda, Cox, Shirane,
  Park, Cross, and Zhong}}]{Polarization}
\bibinfo{author}{\bibfnamefont{B.}~\bibnamefont{Noheda}},
  \bibinfo{author}{\bibfnamefont{D.~E.} \bibnamefont{Cox}},
  \bibinfo{author}{\bibfnamefont{G.}~\bibnamefont{Shirane}},
  \bibinfo{author}{\bibfnamefont{S.-E.} \bibnamefont{Park}},
  \bibinfo{author}{\bibfnamefont{L.~E.} \bibnamefont{Cross}}, \bibnamefont{and}
  \bibinfo{author}{\bibfnamefont{Z.}~\bibnamefont{Zhong}},
  \bibinfo{journal}{Phy. Rev. Lett.} \textbf{\bibinfo{volume}{86}},
  \bibinfo{pages}{3891} (\bibinfo{year}{2001}).

\bibitem[{\citenamefont{Gehring et~al.}(2003)\citenamefont{Gehring, Chen, Ye,
  and Shirane}}]{PMN-10PT}
\bibinfo{author}{\bibfnamefont{P.~M.} \bibnamefont{Gehring}},
  \bibinfo{author}{\bibfnamefont{W.}~\bibnamefont{Chen}},
  \bibinfo{author}{\bibfnamefont{Z.-G.} \bibnamefont{Ye}}, \bibnamefont{and}
  \bibinfo{author}{\bibfnamefont{G.}~\bibnamefont{Shirane}}
  (\bibinfo{year}{2003}), \eprint{cond-mat/0304289}.

\bibitem[{\citenamefont{Xu et~al.}(2003{\natexlab{b}})\citenamefont{Xu,
  Viehland, Li, Shirane, and Gehring}}]{Xu1}
\bibinfo{author}{\bibfnamefont{G.}~\bibnamefont{Xu}},
  \bibinfo{author}{\bibfnamefont{D.}~\bibnamefont{Viehland}},
  \bibinfo{author}{\bibfnamefont{J.~F.} \bibnamefont{Li}},
  \bibinfo{author}{\bibfnamefont{G.}~\bibnamefont{Shirane}}, \bibnamefont{and}
  \bibinfo{author}{\bibfnamefont{P.~M.} \bibnamefont{Gehring}},
  \bibinfo{journal}{Phys. Rev. B} \textbf{\bibinfo{volume}{68}},
  \bibinfo{pages}{212410} (\bibinfo{year}{2003}{\natexlab{b}}).

\bibitem[{\citenamefont{Ye et~al.}(2003)\citenamefont{Ye, Bing, Gao, Bokov,
  Stephens, Noheda, and Shirane}}]{PMN_Ye}
\bibinfo{author}{\bibfnamefont{Z.-G.} \bibnamefont{Ye}},
  \bibinfo{author}{\bibfnamefont{Y.}~\bibnamefont{Bing}},
  \bibinfo{author}{\bibfnamefont{J.}~\bibnamefont{Gao}},
  \bibinfo{author}{\bibfnamefont{A.~A.} \bibnamefont{Bokov}},
  \bibinfo{author}{\bibfnamefont{P.}~\bibnamefont{Stephens}},
  \bibinfo{author}{\bibfnamefont{B.}~\bibnamefont{Noheda}}, \bibnamefont{and}
  \bibinfo{author}{\bibfnamefont{G.}~\bibnamefont{Shirane}},
  \bibinfo{journal}{Phys. Rev. B} \textbf{\bibinfo{volume}{67}},
  \bibinfo{pages}{104104} (\bibinfo{year}{2003}).

\bibitem[{\citenamefont{Burns and Dacol}(1983)}]{Burns}
\bibinfo{author}{\bibfnamefont{G.}~\bibnamefont{Burns}} \bibnamefont{and}
  \bibinfo{author}{\bibfnamefont{F.~H.} \bibnamefont{Dacol}},
  \bibinfo{journal}{Phys. Rev. B} \textbf{\bibinfo{volume}{28}},
  \bibinfo{pages}{2527} (\bibinfo{year}{1983}).

\bibitem[{\citenamefont{Smolensky}(1970)}]{Smolensky}
\bibinfo{author}{\bibfnamefont{G.~A.} \bibnamefont{Smolensky}},
  \bibinfo{journal}{J. Phys. Soc. Jpn.} \textbf{\bibinfo{volume}{28}},
  \bibinfo{pages}{26} (\bibinfo{year}{1970}).

\bibitem[{\citenamefont{Cross}(1987)}]{Cross}
\bibinfo{author}{\bibfnamefont{L.~E.} \bibnamefont{Cross}},
  \bibinfo{journal}{Ferroelectrics} \textbf{\bibinfo{volume}{76}},
  \bibinfo{pages}{241} (\bibinfo{year}{1987}).

\bibitem[{\citenamefont{Hirota et~al.}(2002)\citenamefont{Hirota, Ye, Wakimoto,
  Gehring, and Shirane}}]{PMN_diffuse}
\bibinfo{author}{\bibfnamefont{K.}~\bibnamefont{Hirota}},
  \bibinfo{author}{\bibfnamefont{Z.-G.} \bibnamefont{Ye}},
  \bibinfo{author}{\bibfnamefont{S.}~\bibnamefont{Wakimoto}},
  \bibinfo{author}{\bibfnamefont{P.~M.} \bibnamefont{Gehring}},
  \bibnamefont{and} \bibinfo{author}{\bibfnamefont{G.}~\bibnamefont{Shirane}},
  \bibinfo{journal}{Phys. Rev. B} \textbf{\bibinfo{volume}{65}},
  \bibinfo{pages}{104105} (\bibinfo{year}{2002}).

\bibitem[{\citenamefont{Stock et~al.}(2002)\citenamefont{Stock, Birgeneau,
  Wakimoto, Gardner, Chen, Ye, and Shirane}}]{Stock1}
\bibinfo{author}{\bibfnamefont{C.}~\bibnamefont{Stock}},
  \bibinfo{author}{\bibfnamefont{R.~J.} \bibnamefont{Birgeneau}},
  \bibinfo{author}{\bibfnamefont{S.}~\bibnamefont{Wakimoto}},
  \bibinfo{author}{\bibfnamefont{J.}~\bibnamefont{Gardner}},
  \bibinfo{author}{\bibfnamefont{W.}~\bibnamefont{Chen}},
  \bibinfo{author}{\bibfnamefont{Z.-G.} \bibnamefont{Ye}}, \bibnamefont{and}
  \bibinfo{author}{\bibfnamefont{G.}~\bibnamefont{Shirane}}
  (\bibinfo{year}{2002}), \eprint{cond-mat/0301132}.

\end{thebibliography}
\end{document}